\documentclass[reprint,superscriptaddress,preprintnumbers,amsmath,amssymb,aps,prd,tightenlines,longbibliography]{revtex4-2}

\usepackage{graphicx}
\usepackage{xcolor}
\usepackage[sort&compress]{natbib}
\usepackage{amsmath,amssymb,bm,bbm,slashed,subdepth}
\usepackage{xr-hyper}
\usepackage[colorlinks=true
,urlcolor=blue
,anchorcolor=blue
,citecolor=blue
,filecolor=blue
,linkcolor=red
,menucolor=blue
,linktocpage=true
,pdfproducer=medialab
,pdfa=true
]{hyperref}
\usepackage{cleveref}
\usepackage{enumerate}
\usepackage{epsfig, subfigure}
\usepackage{setspace}
\usepackage{booktabs, tabularx}
\usepackage{units}
\usepackage{placeins}
\usepackage{multirow}
\usepackage{mathtools}
\usepackage[normalem]{ulem}

\newcommand{\erho}{\hat{\mathbf{e}}_\rho}
\newcommand{\ephi}{\hat{\mathbf{e}}_\phi}
\newcommand{\ez}{\hat{\mathbf{e}}_z}
\newcommand{\ex}{\hat{\mathbf{e}}_x}
\newcommand{\ey}{\hat{\mathbf{e}}_y}
\newcommand{\rv}{\mathbf{r}}
\newcommand{\vv}{\mathbf{b}}
\newcommand{\jv}{\mathbf{j}}

\newcommand{\bmax}{B_{\text{max}}}

\newcommand{\be}{\begin{equation}}
\newcommand{\ee}{\end{equation}}
\newcommand{\bea}{\begin{equation}\begin{aligned}}
\newcommand{\eea}{\end{aligned}\end{equation}}
\newcommand{\ga}{g_{a\gamma\gamma}}
\newcommand{\bE}{\mathbf{E}}
\newcommand{\bB}{\mathbf{B}}
\newcommand{\bP}{\mathbf{P}}
\newcommand{\bM}{\mathbf{M}}
\newcommand{\bk}{\mathbf{k}}
\newcommand{\bU}{\mathbf{U}}
\newcommand{\bV}{\mathbf{V}}
\newcommand{\rhoDM}{\rho_{\scriptscriptstyle \textrm{DM}}}
\newcommand{\PBH}{{\scriptscriptstyle \textrm{PBH}}}
\newcommand{\hTT}{h^{\scriptscriptstyle \textrm{TT}}}

\def \i {\mathrm{i}\mkern1mu} 

\begin{document}

\preprint{DESY-22-017}
\preprint{CERN-TH-2022-010}

\title{A novel search for high-frequency gravitational waves with low-mass axion haloscopes}
 

\author{Valerie Domcke}
\affiliation{Theoretical Physics Department, CERN,
1 Esplanade des Particules, CH-1211 Geneva 23, Switzerland}
\affiliation{Institute of Physics, 
\'Ecole Polytechnique F\'ed\'erale de Lausanne (EPFL), CH-1015 Lausanne, Switzerland}

\author{Camilo Garcia-Cely}
\affiliation{Deutsches Elektronen-Synchrotron DESY, Notkestr. 85,
22607 Hamburg, Germany}

\author{Nicholas L. Rodd}
\affiliation{Theoretical Physics Department, CERN,
1 Esplanade des Particules, CH-1211 Geneva 23, Switzerland}

\begin{abstract}
Gravitational waves (GWs) generate oscillating electromagnetic effects in the vicinity of external electric and magnetic fields.
We discuss this phenomenon with a particular focus on reinterpreting the results of axion haloscopes based on lumped-element detectors, which probe GWs in the 100~kHz\,-100~MHz range.
Measurements from ABRACADABRA and SHAFT already place bounds on GWs, although the present strain sensitivity is weak. 
However, we demonstrate that the sensitivity scaling with the volume of such instruments is significant -- faster than for axions -- and so rapid progress will be made in the future.
With no modifications, DMRadio-m$^{\!3}$ will have a GW strain sensitivity of $h \sim 10^{-20}$ at 200~MHz.
A simple modification of the pickup loop used to readout the induced magnetic flux can parametrically enhance the GW sensitivity, particularly at lower frequencies.
\end{abstract}

\maketitle

The present gravitational wave (GW) program is focussed on the nHz to kHz frequency range, motivated by the signals expected from the merging of known compact astrophysical objects.
This focus leaves the ultra-high frequency (UHF) range, above a kHz, largely unexplored, despite its unique opportunity to probe the physics of the very early Universe.
For a recent summary of the challenges and opportunities at high frequencies, see Ref.~\cite{Aggarwal:2020olq}.

\begin{figure}[t]
\includegraphics[width=1.\columnwidth]{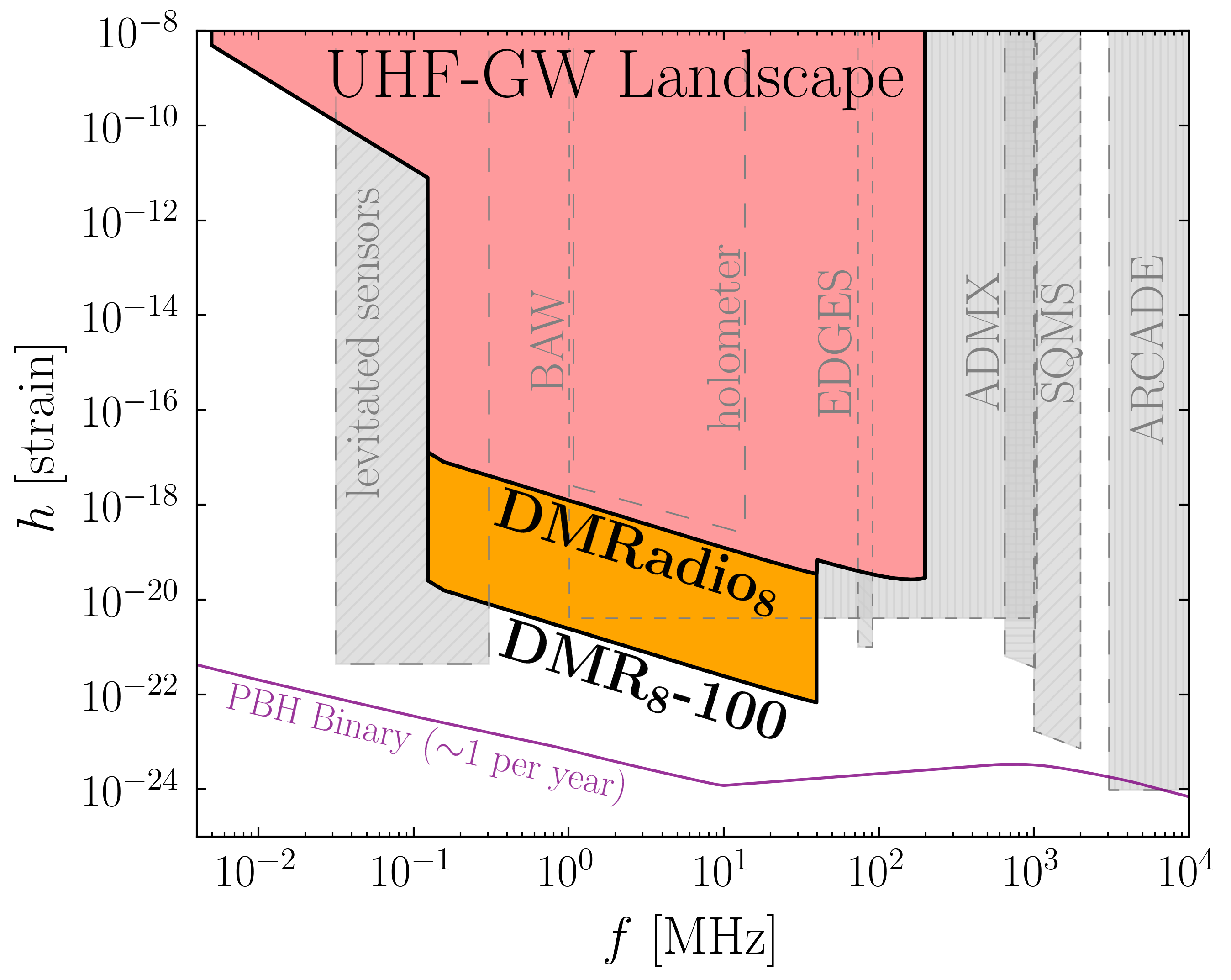}
\vspace{-0.6cm}
\caption{The UHF-GW experimental landscape, with the approach introduced in this work shown in color.
DMRadio$_8$ shows the projected reach of the full suite of DMRadio instruments (50L, m$^{\!3}$, and GUT) adopting our advocated figure-8 pickup loop geometry.
Looking to the far future, we also show the reach of an upscaled DMRadio with a magnetic field volume of 100~m$^{\!3}$, labelled DMR$_8$-100.
A subset of existing proposals in this frequency range are shown in grey, taken from Refs.~\cite{Aggarwal:2020olq,Berlin:2021txa},  as well as an estimate for the required sensitivity to see one signal from primordial black hole (PBH) binaries per year.
Additional specifics are provided in the text.
}
\label{fig:projectedlimits}
\vspace{-0.7cm}
\end{figure}

In this work we propose a novel strategy for the UHF range based on GW electrodynamics: the modified version of electromagnetism applicable in the spacetime metric of a GW.
We show there exists a close analogy to axion electrodynamics -- the appropriate formalism when working in the background of an ultralight axion -- and exploit this connection to convert axion haloscopes into GW telescopes.
In the vicinity of static electric and magnetic fields, a GW sources a small electromagnetic signal oscillating at the GW frequency.
This opens up the possibility of using low-mass axion haloscopes with a frequency range of 100~kHz\,-100~MHz as detectors for UHF GWs.
Our proposed search strategy will utilize the anticipated rapid progress being made in this field by lumped-element axion detectors.
We will show how the existing results of ABRACADABRA~\cite{Kahn:2016aff,Ouellet:2018beu,Ouellet:2019tlz,Salemi:2021gck} and SHAFT~\cite{Gramolin:2020ict} can already be recast as novel limits on GWs.
Going forward, DMRadio~\cite{Chaudhuri:2014dla,Silva-Feaver:2016qhh,SnowmassOuellet,SnowmassChaudhuri} will dramatically extend both the axion and GW reach at these frequencies.
As we will demonstrate, the reach of DMRadio can be enhanced with a simple modification to the signal readout, by using a semicircular ``figure-8'' loop to measure the magnetic flux.
With this adjustment, the future reach of DMRadio, as shown in Fig.~\ref{fig:projectedlimits}, will represent both a competitive and complementary approach to UHF GWs.
Looking even further into the future, the scaling of the GW reach with the instrument volume is particularly advantageous, which we highlight with the larger DMR$_8$-100.

Figure~\ref{fig:projectedlimits} also depicts existing proposals for the UHF band, including optically levitated sensors~\cite{Arvanitaki:2012cn,Aggarwal:2020umq}, bulk acoustic wave (BAW) devices~\cite{Goryachev:2014yra} (see also Ref.~\cite{Goryachev:2021zzn}), interferometers such as the holometer~\cite{Holometer:2016qoh,Martinez:2020cdh,Vermeulen:2020djm}, current and future microwave cavity instruments, such as the axion haloscope approach introduced in Ref.~\cite{Berlin:2021txa} for ADMX~\cite{ADMX:2021nhd,ADMX:2019uok,ADMX:2018ogs} and SQMS (see also Ref.~\cite{Herman:2020wao}), as well as cosmological probes of GWs~\cite{Domcke:2020yzq} based on observations by the radio telescopes EDGES and ARCADE~\cite{Bowman:2018yin,Fixsen_2011}.
We emphasize that the different proposals we show are at varied levels of maturity, and therefore caution against overly quantitative comparisons.
Instead, we refer to the specific references for details. 
Our proposal is related to the (inverse) Gertsenshtein effect~\cite{Gertsenshtein,Boccaletti}, which describes the conversion of GWs into photons, a concept that has been proposed as method to search for GWs, in different frequency regimes, in the laboratory~\cite{Boccaletti,Zeldovich,DeLogi:1977qe,Ejlli:2019bqj} and cosmology~\cite{Raffelt:1987im,PhysRevLett.74.634,Dolgov:2012be,Fujita:2020rdx,Pshirkov:2009sf,Ringwald:2020ist}. 
For electromagnetic GW detectors in a broader sense, see also pioneering work in Refs.~\cite{Braginskii:1973vm,Grishchuk:1975tg,Caves:1979kq,Pegoraro:1978gv,Pegoraro:1977uv,Reece:1984gv,Reece:1982sc}.

We organize the discussion as follows.
We begin with several general comments on the modification to electrodynamics induced by a passing GW, before specializing to the case of interest: the sensitivity of instruments that use a toroidal magnetic field.
We then outline how we can exploit this to recast existing ABRA and SHAFT results, and future DMRadio searches.
In the Supplementary Material (SM) we provide the full details of our calculations and a brief discussion of GW sources in the UHF band.

\noindent
{\bf Gravitational Wave Electrodynamics.}
%
We describe the spacetime in the presence of a GW by the linearized metric $g_{\mu \nu} = \eta_{\mu \nu} + h_{\mu \nu}$, with $|h_{\mu \nu}| \ll 1$.
The perturbation to the flat-space metric generates a correction to the kinetic term of electromagnetism, which can be written as the following effective current,
\be
\partial_\nu F^{\mu\nu}= j^\mu_\text{eff}= \left(-\nabla\cdot \bP , \,  \nabla \times \bM + \partial_t \bP \right)\!.
\label{eq:PM0}
\ee
Here $F^{\mu \nu}$ is the electromagnetic field-strength tensor, and as we demonstrate in the SM, the \emph{effective} polarization and magnetization vectors are
\bea
P_i &=-h_{i j} E_{j}+\tfrac{1}{2} h E_{i}+h_{00} E_{i}-\epsilon_{i j k} h_{0 j} B_{k}, \\ 
M_i &=-h_{i j} B_{j}-\tfrac{1}{2} h B_{i}+h_{j j} B_{i}+\epsilon_{i j k} h_{0 j} E_{k},
\label{eq:PM}
\eea
where $h = {h^{\mu}}_{\!\mu}$.
Manifestly, near large external electric or magnetic fields, GWs will source oscillating fields, the detection of which is the focus of this work.

The above formalism facilitates a comparison with axion electrodynamics.
In particular, the coupling between the axion, $a$, and electromagnetism is also described by Eq.~\eqref{eq:PM0}, but with $\bP = \ga a\,\bB$ and $\bM = \ga a\, \bE$, as follows from ${\cal L} \supset \ga a\, \bE \cdot \bB$~\cite{McAllister:2018ndu,Tobar:2018arx,Ouellet:2018nfr}, where $ \ga$ is the axion-photon coupling.

We can extend this analogy in order to estimate the expected GW sensitivity of axion haloscopes.
For both the axion and GW, the magnitude of the induced fields is controlled by a dimensionless combination, either the strain $h \sim |h_{\mu \nu}|$ or $\ga a$.
The axion dark-matter program aims to probe the QCD axion, for which $m_a f_a \sim m_\pi f_\pi$, where $f_a$ is the axion decay constant, with $\ga = \alpha/2\pi f_a$.
Matching the strain sensitivity to the average $\ga a~\sim \ga \sqrt{\rhoDM}/m_a $ for the QCD axion, we find $h \sim \alpha \sqrt{\rhoDM}/2\pi \, m_\pi f_\pi\sim 10^{-22}$.
This estimate sets the scale for the GWs that can be potentially detected (see Fig.~\ref{fig:projectedlimits}).
In this argument we introduced the electromagnetic fine structure constant $\alpha$, the local dark-matter density $\rhoDM$, as well as the pion mass and decay constant, $m_{\pi}$ and $f_{\pi}$.
Cosmological GW sources, which are typically isotropic and incoherent, are bounded by constraints on the total amount of radiation in the Universe to satisfy $h \lesssim 10^{-29} \; (100~\text{MHz}/f) \; \Delta N_\text{eff}^{1/2}$, well below our estimated reach. 
More promising search targets are rare exotic astrophysical GW sources.
As a concrete example, if primordial black holes (PBH) with $m_\PBH \ll M_{\odot}$ contribute to the dark-matter density, then some fraction of these will exist in binaries and emit high frequency GWs through their inspiral and eventual merger phase.
In order to estimate the size of this signal, we take the most up-to-date estimates of the fraction of PBHs in binaries and their expected merger rate (see Sec. 4.6 of Ref.~\cite{Franciolini:2021nvv} for a recent review, although we emphasize that uncertainties remain such as the impact of accretion on the merger rate).
Combining the merger rate with the assumption that PBHs saturate the relevant microlensing constraints~\cite{Carr:2021bzv}, and accounting for the local overdensity of binaries given by the Milky Way halo, we arrive at the estimated sensitivity shown in Fig.~\ref{fig:projectedlimits} required in order to see one event per year at that frequency (marginalizing over $m_\PBH$).
See the SM for further details of this calculation and a discussion of other putative signals.
We thus primarily focus on localized, approximately coherent GW signals in the following.

In the transverse-traceless (TT) gauge, the non-vanishing components of a plane GW read 
\begin{eqnarray}
\hTT_{ij}&\!\!=& \! \! \left[ \left({\rm U}_i {\rm U}_j\!-\!{\rm V}_i {\rm V}_j\right)\! h^+
\!+\! \left({\rm U}_i {\rm V}_j\!+\!{\rm V}_i {\rm U}_j\right)\! h^\times
\right]  \!\frac{e^{\i(\mathbf{k}\cdot \rv - \omega t)}}{\sqrt{2}},\nonumber\\
\hat{\bk}&\!\!=& s_{\theta_h} \erho^{\phi_h}+c_{\theta_h}\ez,\,\,
\bV = \ephi^{\phi_h},\,\,
\bU = \mathbf{V} \times \hat{\bk},
\label{eq:planewave}
\end{eqnarray}
where $\phi_h$ and $\theta_h$ are, respectively, the azimuthal and inclination angles of the GW, and throughout we employ the shorthand $s_\alpha=\sin \alpha$ and $c_\alpha = \cos \alpha$.
$h^+$ and $h^{\times}$ are the strain amplitude associated with the plus and cross polarizations.
The choice of $\bV$ is a convention, any unit vector perpendicular to $\bk$ can be adopted.
We note that different choices will mix the definitions of $h^+$ and $h^{\times}$.

In the TT gauge, the form of Eq.~\eqref{eq:planewave} appears to considerably simplify Eq.~\eqref{eq:PM}, although for our purposes this can be deceptive because the experimentally generated electric and magnetic fields in the equation are not naturally defined in the TT frame.
Instead, throughout this work we will operate exclusively in the frame where all detector quantities are defined. This is the proper detector frame, in which following Ref.~\cite{Berlin:2021txa} (see also Refs.~\cite{FortiniGualdi,Marzlin:1994ia,Rakhmanov:2014noa}), we find
\begin{eqnarray}
h_{00}\! &=&  \omega^2  F( \bk \cdot \rv) \,\vv \cdot \rv , \hspace{20pt}b_j \equiv r_i \hTT_{ij}\big|_{\rv=0} , \nonumber \\
h_{0i}\! &=& \frac{1}{2} \omega^2 \left[F(\bk \cdot \rv)-\i F'(\bk \cdot \rv) \right] \!\left(\hat{\mathbf{k}} \cdot \rv  \,\,b_i - \vv \cdot \rv  \,\, \hat{k}_i \right)\!, \label{eq:hDF} \\ 
h_{ij}\! &=& - \i \omega^2  F'(\bk \cdot \rv) \left( |\rv|^2 \, \hTT_{ij}\big|_{\rv=0} +\vv \cdot \rv \, \delta_{ij} \!-\! b_i  r_j \!-\! b_j r_i\right)\!,\nonumber
\end{eqnarray}
with $F(\xi)= (e^{\i \xi } -1-\i \xi)/\xi ^2 \approx -1/2+{\cal O}(\xi)$. See the SM for details.
On very general grounds, Eq.~\eqref{eq:hDF} shows that the effective current and therefore the fields it induces are rapidly suppressed for frequencies below the inverse length scale of the instrument.
As we show, there are further suppressions if the experiment is highly symmetric.

\begin{figure}[t]
\hspace{-0.3cm}\includegraphics[width=0.8\columnwidth]{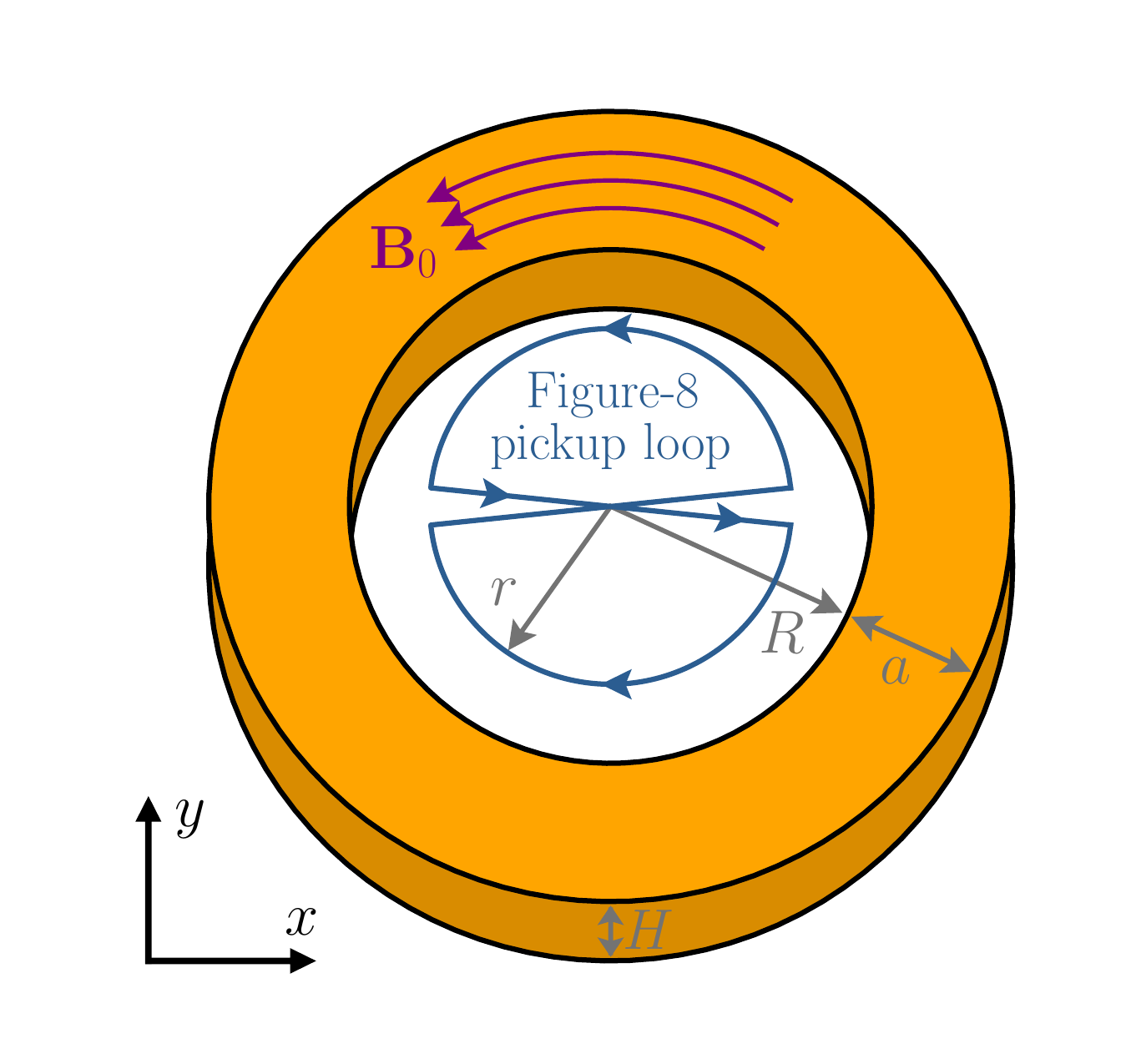}
\vspace{-0.6cm}
\caption{The detector geometry we consider for the detection of GWs.
The experimental setup follows ABRA~\cite{Kahn:2016aff}: a toroidal magnetic field $\bB_0$, as given in Eq.~\eqref{eq:B0}, is generated inside a toroid of inner radius $R$, width $a$, and height $H$.
In the presence of a GW or axion, a magnetic flux is generated in a pickup loop placed at the center of the toroid, where $\bB_0=0$.
Axion detection makes use of a circular pickup loop of radius $r$, whereas for optimal GW detection we advocate the figure-8 configuration depicted, made of two oppositely oriented semicircles.
}
\label{fig:detector}
\end{figure}

\noindent
{\bf Application to a Toroidal Magnetic Field.}
%
We consider now an explicit experimental configuration to detect the oscillating fields sourced by a passing GW.
In particular, we follow the original ABRA proposal~\cite{Kahn:2016aff} of establishing a DC toroidal magnetic field,
\begin{align}
\bB_0 = \bmax (R/\rho) \, \ephi,&& \text{for}\,\, R<\rho<R+a,
\label{eq:B0}
\end{align}
and zero otherwise.
The corresponding geometry is depicted in Fig.~\ref{fig:detector}.
The combined effect of the magnetic field and a GW is the effective current in Eq.~\eqref{eq:PM0} that, according to the Biot-Savart law, sources a magnetic field in the region $\rho < R$ such that
\be
B_z(\rv') = \int_{\text{toroid}} \hspace{-0.6cm} \mathrm{d}^3 \rv\, \frac{(j_{\rho} \ephi - j_{\phi} \erho) \cdot (\rv'-\rv)}{4\pi\,|\rv'-\rv|^3},
\label{eq:Bz}
\ee
where $j_{\rho} = \jv_\text{eff} \cdot \erho$ and $j_{\phi} = \jv_\text{eff} \cdot \ephi$.
As we review in the SM, corrections to the Biot-Savart law from the displacement current enter only at ${\cal O}(\omega^4)$.
To detect this magnetic field, a pickup loop is placed at the center of the toroid, which will be sensitive to a magnetic flux equal to Eq.~\eqref{eq:Bz} integrated over the area of the loop.
We next consider two different loop geometries, beginning with the approach used in existing axion instruments.

\textit{Circular pickup loop:}
%
For a complete circle, the $j_{\rho}$ contribution vanishes -- independent of the form of $j^\mu_\text{eff}$ -- leaving
\be
\Phi =  -\int_{\text{loop}} \hspace{-0.4cm} \mathrm{d}^2 \rv'\,\int_{\text{toroid}} \hspace{-0.6cm} \mathrm{d}^3 \rv\, \frac{j_{\phi} \erho \cdot (\rv'-\rv)}{4\pi\,|\rv'-\rv|^3}.
\ee
A shift of $\phi'\to\phi'+\phi$ removes the $\phi$ dependence in the integrand in all terms except $j_{\phi}$.
Using the results in the previous section, we obtain
\be
\int_0^{2\pi}\!\!\mathrm{d}\phi\, j_\phi =  \i\sqrt{2}  \pi\omega h^\times \bmax\frac{R}{\rho} \,J_2 (\omega \rho s_{\theta_h})  e^{\i \omega (z c_{\theta_h}-t)}.
\label{eq:jphiint}
\ee
Independent of the incident GW direction, the $h^+$ component decouples.
Furthermore, in contrast to what would be naively expected from Eq.~\eqref{eq:hDF}, the flux receives no contribution at ${\cal O }(\omega^2)$ because the Bessel function of second order satisfies  $J_2(x)= x^2/8+{\cal O}(x^3)$.
Instead, in the limit $R \ll H \ll 1/\omega$ the leading contribution to the flux is 
\be
\Phi
= \frac{\i\, e^{-\i \omega t}}{16\sqrt{2}}\, h^{\times} \omega^3 \bmax \pi r^2 R a (a+2R) s^2_{\theta_h}.
\label{eq:w3Phi}
\ee
Nonetheless, a GW will generate a flux, and existing axion searches for such a flux can constrain UHF-GWs.
In this regard, Eq.~\eqref{eq:w3Phi} may be compared against the equivalent quantity for axions~\cite{Kahn:2016aff}, for which $j_{\rho}=0$, $j_{\phi} = \ga (\partial_t a) \bmax R/\rho$, and expanding in $R/H$ yields
\be
\Phi = e^{-\i \omega t}\, \ga \sqrt{2\rhoDM} \bmax \pi r^2 R \ln (1+a/R).
\label{eq:axionPhi}
\ee
To isolate the fate of the expected $\omega^2$ contributions, let us note that, at leading order in $\omega$, 
\begin{eqnarray}
j_\phi &=&  \frac{\omega^2 B_\text{max} R\, e^{-\i\omega t}}{3 \sqrt{2} \rho} \left[ h^\times \left( \rho \, c_{\phi} s_{\theta_h} + z c_{2\phi} c_{\theta_h} \right)  \right.  \label{eq:js} \\
& & \hspace{2cm}  \left. - h^+ s_\phi \left(\rho \, c_{\theta_h} s_{\theta_h} + z (1 + c_{\theta_h}^2) c_{\phi}\right) \right]\!, \nonumber \\
j_\rho &=& \frac{\omega^2 B_\text{max} R \, e^{-\i\omega t}}{6 \sqrt{2} \rho} \left[ h^\times s_{\phi} \left( 8 z c_{\theta_h} c_{\phi} - \rho \, s_{\theta_h} \right) \right. \nonumber \\
& & \left.\hspace{0.5cm}- h^+ \left(\rho \, c_{\theta_h} s_{\theta_h} c_{\phi} + z (4 s_{\phi}^2 - 4 c_{\phi}^2 c_{\theta_h}^2 - 5 s_{\theta_h}^2 ) \right)\right]\!, \nonumber 
\end{eqnarray}
where we take $\phi_h=0$.
(It can be restored by replacing $\phi\to \phi-\phi_h$.)
All of these terms will vanish for a circular pickup loop geometry: as noted above $j_{\rho}$ cannot contribute in this case in general, and from the explicit expression we see that at ${\cal O}(\omega^2)$, $\int^{2\pi}_0 \mathrm{d}\phi\, j_\phi=0$.
Having identified this, however, we can see that with an alternative geometry, these leading terms will survive.
To be explicit, using the above currents we can compute the leading contribution to the magnetic field in the plane at the vertical center of the toroid, from Eq.~\eqref{eq:Bz},
\bea
B_z&(\rho',\phi') = \frac{e^{-\i \omega t}}{4\sqrt{2}}\, \omega^2 B_\text{max} \rho' R \ln (1+a/R) s_{\theta_h} \\
&\hspace{0.5cm}\times \left( h^{\times} \cos (\phi'-\phi_h) - h^+ c_{\theta_h} \sin (\phi'-\phi_h) \right)\!.
\label{eq:leadingBz}
\eea
The sinusoidal variation of both polarizations demonstrates that the maximum flux is achieved with a pickup loop with oppositely oriented semicircles for $\phi' \in [0,\pi)$ and $\phi' \in [\pi,2\pi)$---the ``figure-8'' shown in Fig.~\ref{fig:detector}.

\textit{The ``figure-8'' configuration:}
%
Integrating Eq.~\eqref{eq:leadingBz} over the figure-8 configuration yields
\bea
\Phi_8 =\,& \frac{e^{-\i \omega t}}{3\sqrt{2}}\, \omega^2 \bmax r^3 R \ln \left(1+a/R\right) s_{\theta_h} \\
& \times \left(h^\times s_{\phi_h} -h^+c_{\theta_h} c_{\phi_h}\right)\!.
\label{eq:w2Phi}
\eea
The result is now ${\cal O}(\omega^2)$ and sensitive to both polarizations.
While it will maximize the GW sensitivity, the figure-8 pickup loop is insensitive to the axion signal, as the $B_z$ generated by the latter is independent of $\phi'$.
A single semicircle is sensitive to both, with fluxes given by half of Eqs.~\eqref{eq:axionPhi} and \eqref{eq:w2Phi}, respectively.
Comparing the two expressions, we also see that to compensate the $\omega^2$ factor, the GW flux scales with an extra power of $r$, indicating the improved volume scaling over the axion flux.
Accounting for the minimal inductance of the pickup loop~\cite{Kahn:2016aff}, we see $\Phi_8 \propto V^{7/6}$ whereas $\Phi_a \propto V^{5/6}$.
Ultimately, the beneficial volume scaling can be traced back to the fact that our measurement is linear in the induced fields, which themselves are proportional to $h$ for the GW or $a$ for the axion.

From Eq.~\eqref{eq:w2Phi}, we note that the response to the two polarizations differs and depends on the direction of the incoming GW.
With two identical detectors angled appropriately, polarization measurements as well as sky localization become possible (see also Ref.~\cite{Foster:2020fln,Chen:2021bdr}).
For a sufficiently coherent source, one may hope to use the Earth's rotation or even a mechanical rotation of the experimental setup to break these degeneracies with a single detector.

\noindent
{\bf Gravitational Wave Sensitivity.}
%
Low-mass axion haloscopes perform a search for anomalous magnetic flux, and in the absence of a significant signal above background, interpret the results as limits on $\ga$ through the use of Eq.~\eqref{eq:axionPhi}.
With the same equation we can convert existing and projected limits on $\ga$ into limits on $\Phi_a$, which we can recast as strain sensitivities when compared with our predictions for the GW flux.
The procedure is not quite as simple as equating the latter to $\Phi_a$, however.
The sensitivity also depends on the relative coherence time of the two signals---a longer coherence time corresponds to a narrower signal in the frequency domain, which in general can be more sensitively detected over the background (for an extended discussion, see Ref.~\cite{Dror:2021nyr}).
The coherence time of a signal with mean frequency $\bar{\omega}$ is $\tau \sim 2\pi Q/\bar{\omega}$, where $Q$ is the quality of the signal, a dimensionless measure of the inverse width of the frequency distribution.
The non-relativistic nature of axion dark-matter implies a highly coherent signal with $Q_a \sim 10^6$, such that $\tau_a \sim (1~\text{neV}/m_a)~\mu\text{s}$.
Considering experimental runtimes longer than $\tau$, then the flux sensitivity will scale as $\Phi \propto Q^{1/4}$~\cite{Budker:2013hfa}, and so our actual limit on the GW flux is given by $\Phi_a (Q_a/Q_h)^{1/4}$.
Beyond this we assume the signal persists during the relevant experimental runtime, and fix the incident GW direction as $\hat{\bk} = \ey$.

\begin{figure}[!t]
\label{fig:circular}
\includegraphics[width=1.\columnwidth]{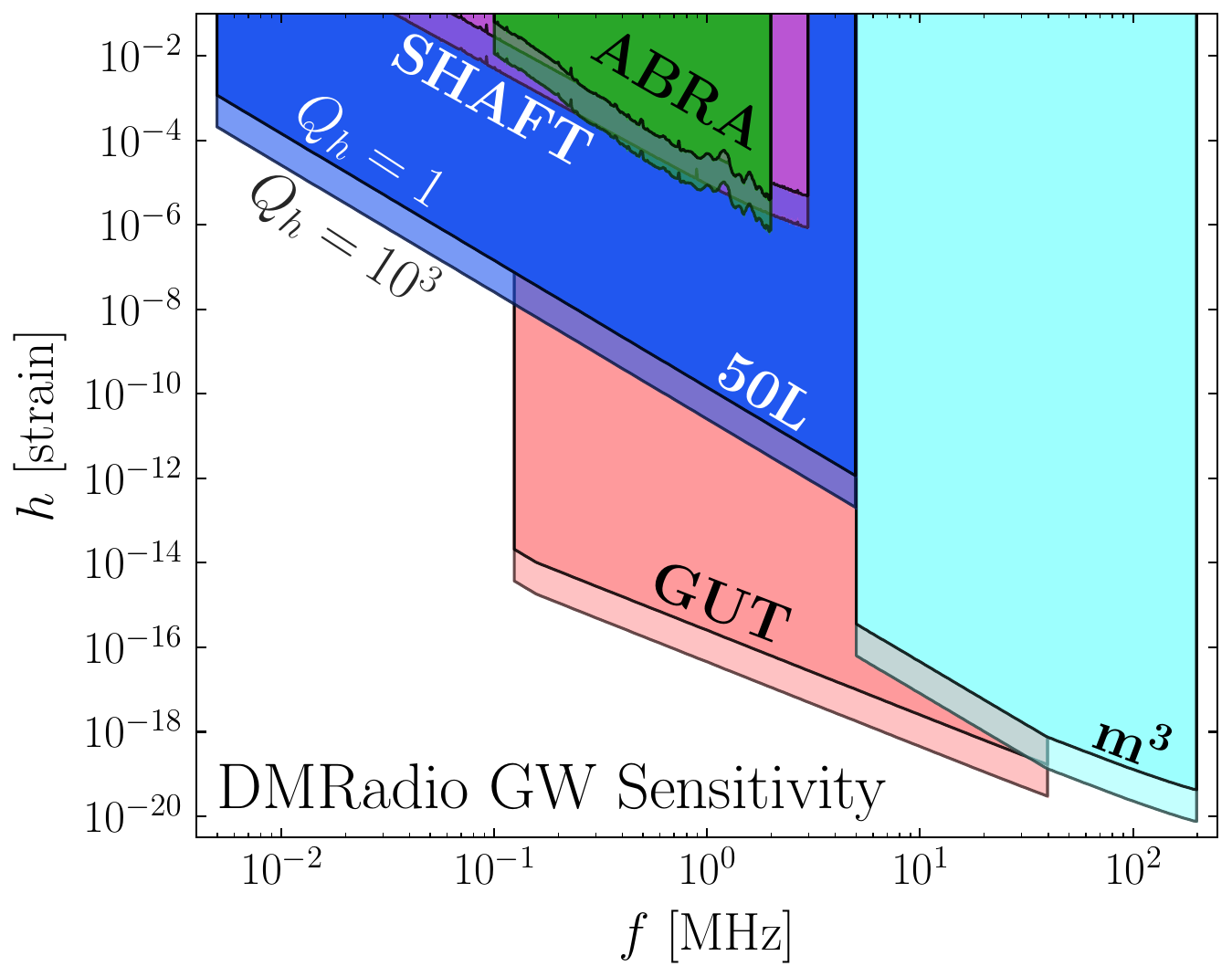}
\vspace{-0.6cm}
\caption{The GW strain sensitivity of low-mass axion haloscopes.
We recast the existing limits obtained by ABRA~\cite{Salemi:2021gck} (green) and SHAFT~\cite{Gramolin:2020ict} (purple). For DMRadio we use the projected future sensitivity of the three instruments that will make up that program: 50L (blue), $m^{\!3}$ (cyan), and GUT (pink)~\cite{SnowmassOuellet,SnowmassChaudhuri}.
In each case, results are shown for two choices of the GW signal coherence, $Q_h = 1$ (opaque) and $Q_h = 10^3$ (transparent).
All results assume a circular pickup loop, for results using the optimal figure-8, see Fig.~\ref{fig:projectedlimits}.
}
\label{fig:rescale}
\end{figure}

What remains is to determine a value for $Q_h$.
This will depend on the specific source.
As mentioned already, the localized sources that are our focus can be coherent, but are not expected to be as extremely coherent as a dark-matter signal (see the SM for further discussion).
As benchmarks, we therefore consider sensitivity to both a coherent $Q_h = 10^3$ and incoherent $Q_h = 1$ signal, to indicate the dependence upon this choice.

With these choices, in Fig.~\ref{fig:rescale} we show the reach of existing and future haloscopes, taking the default circular pickup loop geometry.
This roughly amounts to using Eqs.~\eqref{eq:w3Phi} and \eqref{eq:axionPhi}, except that in all figures we use the full flux expressions (rather than these leading order results), which are provided in the SM.
For ABRA-10~cm~\cite{Salemi:2021gck} and SHAFT~\cite{Gramolin:2020ict}, we recompute the GW flux for the explicit geometries specified in those works, as well as accounting for the ferromagnetic core adopted by SHAFT.
The future projections of DMRadio-50L, m$^{\!3}$, and GUT are obtained by scaling up the toroidal geometry of the ABRA-10~cm instrument to the volumes of these future experiments specified in Refs.~\cite{SnowmassOuellet,SnowmassChaudhuri}.
As the exact parameters of these instruments have yet to be specified, these results should be interpreted as representative of the parametric reach---the sensitivities will change by ${\cal O}(1)$ amounts once the geometry of the instruments is known.
For instance, while the 50L instrument will adopt a toroidal geometry, m$^{\!3}$ will be solenoidal, and the geometry of DMRadio-GUT has not yet been specified.
In all cases, there is a suppression in the sensitivity at lower frequency, which is a result of the $\omega^3$ factor in Eq.~\eqref{eq:w3Phi}.

To overcome this, in Fig.~\ref{fig:projectedlimits} we show the combined sensitivity if the DMRadio instruments adopted the figure-8 configuration (assuming $Q_h=10^3$).
While such DMRadio results are still more than a decade away, in order to highlight the significant volume scaling of our approach, we also show the reach of a scaled up version of DMRadio-GUT.
In particular, DMR$_8$-100 is an instrument with the same toroidal geometry as above, but scaled up to a magnetic field volume of 100~m$^{\!3}$, the largest instrument suggested in the original ABRA proposal~\cite{Kahn:2016aff}.
(The magnetic field volume, $V_B$, is also referred to as the effective volume~\cite{Gramolin:2020ict} or denoted by ${\cal G} V$, where ${\cal G}$ is the geometric coupling $V$ is the volume of the toroid~\cite{Ouellet:2018beu,Ouellet:2019tlz,Salemi:2021gck}.)

\noindent
{\bf Conclusions.}
%
We provide a formulation of GW electrodynamics which demonstrates that low-mass axion haloscopes are also UHF-GW telescopes.
Through the use of an optimized figure-8 pickup loop geometry, the DMRadio program may discover not only the dark matter of our Universe, but also exotic sources of GWs.

Going forward, there are several questions opened by our results that warrant further study.
Determining the optimal experimental configuration that allows DMRadio to unlock the $\omega^2$ GW scaling, while maintaining full axion sensitivity will be critical.
To that end, including both a circular and figure-8 loop in the detector may be required, and we note that the mutual inductance between the two loops would vanish~\footnote{We thank Jon Ouellet for this observation.}.
Further, lumped-element instruments have at present only performed measurements for $\omega \ll 1/R$.
Given the $\omega^2$ scaling of the GW effective current, experimental results at higher frequencies would be particularly welcome.
From a theoretical perspective, it would also be interesting to understand the limit where $\omega \sim 1/R$, where a calculation based on the Biot-Savart law alone is insufficient.
Nevertheless, the exact behavior has yet to be quantified even for axions, and would be particularly interesting to consider for GWs given the $\omega^2$ scaling of the effective current.
Understanding the extension to other geometries will be important, particularly the solenoidal configuration already used by ADMX SLIC~\cite{Crisosto:2019fcj} and that will be adopted by DMRadio-m$^{\!3}$.
Finally, the discussion we presented has been couched purely in terms of the raw GW strain sensitivity.
However, as our understanding of the sources at UHF improves, it will be necessary to reinterpret these results in terms of the specific model parameters describing the source populations.
At such a time, it would also be worthwhile to understand the additional information that multiple instruments could uncover, as we briefly discussed.

While these issues should be resolved, a larger question looms.
Figure~\ref{fig:projectedlimits} highlights that a number of distinct experimental proposals have coalesced on a strain sensitivity of $h \sim 10^{-22}$ for $f \sim$ MHz GWs, a level that is still many order of magnitude away from any signal of the early Universe.
Whether we can hope to probe such strain sensitivities remains to be determined.

{\it Acknowledgements.}
%
We thank Diego Blas, Torsten Bringmann, Sebastian Ellis, Gabriele Franciolini, Yonatan Kahn, Joachim Kopp, Francesco Muia, Jonathan Ouellet, Andreas Ringwald, Chiara Salemi, and Jan Schuette-Engel for helpful discussions.
Additionally, we thank Jonathan Ouellet and the authors of Ref.~\cite{Berlin:2021txa} for comments on the manuscript.
C.G.C is supported by the Deutsche Forschungsgemeinschaft under Germany’s Excellence
Strategy EXC 2121 “Quantum Universe” - 390833306 and by the Alexander von Humboldt Foundation.

{\it Note added.} For an extended discussion of the GW signal from PBH binaries including also the stochastic component see Ref.~\cite{Franciolini:2022xyz}, which presents results obtained in parallel and independently of the results shown here.
Our results are consistent with their findings.

\bibliographystyle{utphys-modified}
\bibliography{ref}

\clearpage
\onecolumngrid
\begin{center}
   \textbf{\large SUPPLEMENTARY MATERIAL \\[.1cm] ``A novel search for high-frequency gravitational \\ waves with low-mass axion haloscopes''}\\[.2cm]
  \vspace{0.05in}
  {Valerie Domcke, Camilo Garcia-Cely, and Nicholas L. Rodd}
\end{center}

\twocolumngrid
\setcounter{equation}{0}
\setcounter{figure}{0}
\setcounter{table}{0}
\setcounter{section}{0}
\setcounter{page}{1}
\makeatletter
\renewcommand{\theequation}{S\arabic{equation}}
\renewcommand{\thefigure}{S\arabic{figure}}
\renewcommand{\thetable}{S\arabic{table}}

\onecolumngrid

In this Supplementary Material we provide detailed derivations of the results quoted in the main text, including somewhat lengthy but useful analytical expressions which can be readily used to analyze more general detector geometries.
We begin in Sec.~\ref{sec:frames} by deriving the expression for the GW in the proper detector frame, Eq.~\eqref{eq:hDF} from the main text, before turning to the heart of our calculations, the effective current induced by a passing GW in Sec.~\ref{sec:current}. 
Finally, in Sec.~\ref{sec:sources} we review possible GW sources in the UHF band.

Throughout this paper, we use Heaviside units with $\hbar=c=1$, $\eta_{\mu\nu} = \text{diag}(- + + +)$, work to leading order in $|h_{\mu \nu}|$, and raise indices with $\eta^{\mu\nu}$.

\section{Gravitational Waves in the Proper Detector Frame}
\label{sec:frames}

The two most commonly used prescriptions for fixing the gauge redundancies in linearized general relativity are the TT gauge and the proper detector frame.
In the former, the coordinate system is defined by freely falling test masses.
The GW tensor takes a particularly simple form obeying
\be
h_{0 \mu} = 0, \quad {h^\mu}_{\!\mu} = 0, \quad \partial^\mu h_{\mu\nu} = 0.
\label{eq:TTgaugecond}
\ee
This simplicity comes at a cost.
In the TT frame, the description of rigid bodies including the various aspects of the detector, become relatively unintuitive, as their coordinates are deformed by a passing GW due to the motion of the coordinate system.
On the other hand, in the proper detector frame the coordinate system is defined by rigid rulers and closely matches the intuitive description of an Earth-based laboratory, with the GW acting as a Newtonian force.
The challenge associated with this frame is that the form of $h_{\mu \nu}$ will be more involved, particularly when the wavelength of the GW is comparable to the size of the detector.
Of the two possible complications, we find the latter preferable, and therefore work in the proper detector frame throughout.
We refer the reader to section 1.1.3 of Ref.~\cite{Maggiore:2007ulw} for a comprehensive discussion of both frames, and to Ref.~\cite{Berlin:2021txa} for specific examples in the context of resonant cavity axion haloscopes.
In this paper, we assume the entire experimental apparatus to be rigid.

The proper detector frame can be implemented with the use of Fermi Normal coordinates~\cite{FortiniGualdi,Marzlin:1994ia,Rakhmanov:2014noa}.
Doing so, the GW tensor is given by
\bea
h_{ij} & = - 2 \sum_{n=0}^\infty \frac{n+1}{(n+3)!} \hat{R}_{ikjl,m_1...m_n} r_k r_l r_{m_1}...r_{m_n}, \\
h_{0i} & = -2 \sum_{n = 0}^\infty \frac{n + 2}{(n+3)!} \hat{R}_{0kil,m_1...m_n} r_k r_l r_{m_1}...r_{m_n}, \\
h_{00} & = -2 \sum_{n = 0}^\infty \frac{n + 3}{(n+3)!} \hat{R}_{0k0l,m_1...m_n} r_k r_l r_{m_1}...r_{m_n}, 
\label{eq:sums}
\eea
with $R_{\mu \nu \rho \sigma}$ denoting the Riemann tensor.
The $m_i$ indices appearing after the comma indicate that a spatial derivative with respect to the direction $m_i$ must be taken.
The notation $\hat{R}$ indicates that the Riemann tensor and its derivatives are to be evaluated at a specific reference point, which we take to be the origin of the coordinate system.
To linear order in $h$, the Riemann tensor is gauge invariant and can thus be evaluated in the particularly simple TT gauge.
Accordingly, we compute Riemann tensor for a GW using $k^{\mu}=(\omega,\,\bk)$ and the conventions we established in \eqref{eq:planewave}. This procedure leads to
\bea
R_{0i0j}  &= \frac{\omega^2}{2} \hTT_{ij},\\
R_{ikjl}  &= \frac{\omega^2}{2} \left(\hat{\bk}_i \hat{\bk}_j \hTT_{kl} + \hat{\bk}_k \hat{\bk}_l \hTT_{ij} - \hat{\bk}_k \hat{\bk}_j \hTT_{il} - \hat{\bk}_i \hat{\bk}_l \hTT_{jk} \right)\!, \\
R_{0ijk}  &= \frac{\omega^2}{2} \left( \hat{\bk}_k \hTT_{ij} - \hat{\bk}_j \hTT_{ik}\right)\!.
\eea
Furthermore, for a GW we have $R_{\mu \nu \rho \sigma,m_1...m_n}  r_{m_1}...r_{m_n} = (\i \bk \cdot \rv)^n R_{\mu \nu \rho \sigma}$.
Using this and the auxiliary function 
\begin{equation}
F(\xi ) = -\sum_{n = 0}^\infty \frac{n + 3}{(n+3)!} (\i\xi)^n = \frac{e^{\i\xi} -1-\i\xi }{\xi^2},
\end{equation}
we can resum the series in Eq.~\eqref{eq:sums}. We find
\begin{align}
h_{00}  =2 F(\bk \cdot  \rv ) \hat R_{0i0j} r_i r_j, 
&&
h_{0i}  =\left[F( \bk \cdot  \rv)-\i F'( \bk \cdot  \rv) \right] \hat R_{0jik} r_j r_k , 
&&
h_{ij}  = -2\, \i \, F'( \bk \cdot  \rv) \hat R_{ikjl} r_k r_l,
\label{eq:h_DF}
\end{align}
which leads to Eq.~\eqref{eq:hDF} in the main text. These expressions agree with those previously shown in Ref.~\cite{Berlin:2021txa} for the particular case $ \hat{\bk} = \ez $  (noting the different sign convention for plane waves adopted in that work).
Corresponding results for arbitrary incident angles were moreover used in the numerical evaluations performed in  Ref.~\cite{Berlin:2021txa}.
From the results above, it is clear that in the proper detector frame, at leading order $h_{\mu \nu} \propto \omega^2$.
As the experimentally measurable flux is linear in $h_{\mu \nu}$, this implies the best scaling we can hope to achieve with frequency is also $\Phi \propto \omega^2$, and indeed this is attained by the figure-8 pickup loop.

\section{The effective current}
\label{sec:current}

The experimentally measurable effect discussed in the main body originates from the effective current a GW generates in the presence of electromagnetic fields.
In this section we first discuss this current in general, introducing the language of effective polarization and magnetization for GW electrodynamics and demonstrating the close analogy with axion electrodynamics.
We then show how a GW generates an effective current, giving explicit analytical expressions for all relevant components.
We conclude this section by turning to the integration of this current, deriving the induced magnetic field and the resulting flux through the pickup loop.

Before we begin, let us be explicit about the order to which we compute our results.
In the main text, we showed analytic expressions to ${\cal O}(\omega^3)$ and expanded assuming $(R+a)/H \ll 1$, in terms of the geometry in Fig.~\ref{fig:detector}.
All sensitivity estimates shown, however, did not expand in the height of the toroid.
Here, using the results of Eq.~\eqref{eq:h_DF} we will compute the effective current to all orders in $\omega$, and with no assumption as to the size of $H$.
However, as we will review below, that we compute the induced magnetic field through the Biot-Savart law implies that we will not be able to determine $B_z$ -- and consequently $\Phi$ -- to ${\cal O}(\omega^4)$ or higher.
As lumped-element axion haloscopes operate mostly in the regime where $\omega \ll 1/R$, this will be sufficient for our purposes; deriving the full result valid even for $\omega \sim 1/R$ would be an interesting future direction.
(We note that, in contrast, the results in Ref.~\cite{Berlin:2021txa} do extend to all orders in $\omega$, as required for resonant cavity instruments.)
Unless otherwise stated, all results outside the main text will not assume $(R+a)/H \ll 1$, and it is those results we used to compute all our projected limits.

\subsection{The Effective Magnetization-Polarization Tensor}

As we will show below, in the absence of ordinary electromagnetic currents,  the effect of axions or GWs on electromagnetic fields can be described by a conserved effective current
\be
\partial_\nu F^{\mu\nu} = j^\mu_\text{eff}.
\label{eq:inhomMax}
\ee
The conservation of this current motivates the introduction of a skew tensor
\be
j^\mu_\text{eff} = \partial_{\nu} {\cal M}_\text{eff}^{\nu \mu},
\label{eq:Meff}
\ee
where ${\cal M}_\text{eff}^{\nu \mu}$ is the effective magnetization-polarization tensor.
Its six components can be expressed in terms of two vectors,
\be
P^i = {\cal M}_\text{eff}^{0 i}, \quad
M_i = - \frac{1}{2} \epsilon_{ijk} {\cal M}_\text{eff}^{jk}.
\ee
These are the effective polarization and magnetization, respectively.
In terms of these two vectors, the effective current then becomes
\be
j^\mu_\text{eff} = (- \nabla \cdot \bP,\, \nabla \times \bM + \partial_t \bP),
\ee
as in Eq.~\eqref{eq:PM0}.
From this, we can identify an effective charge $\rho_\text{eff} = - \nabla \cdot \mathbf{P}$ and an effective 3-current $\jv_\text{eff} = \nabla \times \bM + \partial_t \bP$.
Maxwell's equations take the familiar form
\bea
\nabla \cdot \bE &= - \nabla \cdot \mathbf{P}, \\
\nabla \cdot \bB &= 0, \\
\nabla \times \bE &= - \partial_t \bB, \\
\nabla \times \bB &= \partial_t \bE  + \nabla \times \bM + \partial_t \bP.
\label{eq:MaxwellPM}
\eea
Both axions and GWs can be formally conceived as a continuous medium~\cite{McAllister:2018ndu,Tobar:2018arx,Ouellet:2018nfr,landau1975classical}, as the expressions written above make clear.
Before turning to GWs, let us briefly review the explicit case of axion electrodynamics.
There, the effective current is given by
\be
j^\mu_\text{eff} = \partial_{\nu} \left( \ga a \tilde{F}^{\nu \mu} \right),
\ee
where the derivative will only act non-trivially on the axion due to the Gauss-Faraday law, $\partial_\nu\tilde{F}^{\mu\nu}=0$.
Comparing with Eq.~\eqref{eq:Meff}, we see ${\cal M}_\text{eff}^{\mu \nu} = \ga a \tilde{F}^{\mu \nu}$, so that~\cite{McAllister:2018ndu,Tobar:2018arx,Ouellet:2018nfr}
\be
\bP = \ga a \bB, \quad
\bM = \ga a \bE.
\ee
Substituting these into Eq.~\eqref{eq:MaxwellPM} recovers the equations of axion electrodynamics.
If the axion field is non-relativistic, such that $|\partial_t a| \gg |\nabla a|$, then the leading contribution arises from the effective 3-current $\jv_\text{eff} \simeq \partial_t \bP = \ga \partial_t (a \bB)$, which is the effect that underpins much of the axion dark-matter program.
For relativistic axions, however, all contributions are relevant and the equations match those considered in Ref.~\cite{Dror:2021nyr}.

\subsection{Gravitational Waves as a Source of the Effective Current}

We will show now that, in a GW background and in absence of ordinary charges, Maxwell's equations can be written in the form of Eq.~\eqref{eq:MaxwellPM}. 
In curved space-time, the homogenous Maxwell equations read~\cite{landau1975classical}
\bea
0 &=\nabla_{\mu} F_{\nu \rho}+\nabla_{\nu} F_{\rho \mu}+\nabla_{\rho} F_{\mu \nu}
\\
&=\partial_{\mu} F_{\nu \rho}+\partial_{\nu} F_{\rho \mu}+\partial_{\rho} F_{\mu \nu},
\label{eq:convarianthom}
\eea
whose solution is  $F_{\alpha\beta} =\partial_\alpha A_\beta-\partial_\beta A_\alpha$, independent of the background metric. This demonstrates that the homogeneous Maxwell's equations are not affected by the presence of the GW.
On the other hand, their inhomogeneous counterparts become
\be
\nabla_\nu \left(  g^{\alpha \mu}  F_{\alpha\beta} g^{\beta \nu}\right) = j^\mu,
\label{eq:covariantinhom}
\ee
where $j^\mu$ describes conventional charges and currents.
Using the properties of the divergence, this can be written as 
\be
\partial_\nu \left( \sqrt{-g}\,  g^{\alpha \mu}  F_{\alpha\beta}\,  g^{\beta \nu}\right)  = \sqrt{-g} \, j^\mu.
\label{eq:Maxwell2}
\ee
The smallness of $|h_{\mu\nu}|$ implies that all expressions can be expanded perturbatively.
To first order in $h$, we have $g^{\alpha \mu} F_{\alpha\beta} \, g^{\beta \nu} \simeq   F^{\mu\nu}- F_{\alpha}^{\,\,\nu} h^{\alpha\mu} -{F^{\mu}}_{\!\beta } h^{\beta\nu}$, as well as $\sqrt{-g} \simeq 1+h/2$. 
Hence
\bea
\partial_\nu\, \left( \left(1+\frac{h}{2}\right) F^{\mu\nu}- F_{\alpha}^{\,\,\nu} h^{\alpha\mu} - {F^{\mu}}_{\!\beta } h^{\beta\nu} \right) = \left(1+\frac{h}{2}\right)\, j^\mu +{\cal O} (h^2), 
\eea
which leads to 
\be
\partial_\nu F^{\mu\nu} =\left(1 + \frac{1}{2} h\right) j^{\mu}
+\partial_\nu \left(-\frac{1}{2} h \, F^{\mu\nu} + F_{\alpha}^{\,\,\nu} h^{\alpha\mu} +{F^{\mu}}_{\!\beta} h^{\beta\nu}  \right) +{\cal O} (h^2).
\label{eq:Maxwell3}
\ee
In regions where $j^{\mu}=0$, the source term on the right-hand side can be written as an effective current induced by the GW
\be
j_\text{eff}^{\mu} = \partial_{\nu} \left( -\frac{1}{2} h \, F^{\mu\nu} +F^{\mu \alpha} {h^{\nu}}_{\!\alpha} - F^{\nu\alpha} {h^{\mu}}_{\!\alpha} \right)\!.
\label{eq:jeffGW}
\ee
Conceptualizing the effect of the GW as an effective current was discussed in Ref.~\cite{Herman:2020wao}, working in the TT gauge for a uniform field.
The extension to the proper detector frame and the importance thereof was pointed out in Ref.~\cite{Berlin:2021txa}.

Note that, while there will be ${\cal O}(h)$ corrections to the electromagnetic fields -- indeed, these are the fields we aim to search for experimentally -- these do not enter Eq.~\eqref{eq:jeffGW}, as such terms would be ${\cal O}(h^2)$.
Instead, the fields that enter are those that were established by the experimental apparatus.
To keep track of this, one can introduce a bookkeeping notation $F^{\mu \nu} = F^{\mu \nu}_{0} + F^{\mu \nu}_{h} + {\cal O}(h^2)$, in which case only $F^{\mu \nu}_{0}$ enters the effective current.
(A similar notation is often employed in the axion literature, where the expansion parameter is $\ga$, see e.g. Refs.~\cite{Ouellet:2018nfr,Dror:2021nyr}.)
We will occasionally make use of this notation, but where we do not, the relevant order of terms can always be determined from context.

Furthermore, in the presence of ordinary currents, care must be taken.
As follows from Eq.~\eqref{eq:Maxwell3}, unless $h=0$, one cannot simply add $j^\mu$ and the effective current in order to account for GW effects.
In fact, while $\partial_\mu j_\text{eff}^{\mu}=0$, for ordinary currents we have instead, $\partial_\mu \left((1+h/2)j^\mu\right) =0$, as a consequence of charge conservation in an arbitrary space-time, i.e  $0=\sqrt{-g} \,\nabla_\mu j^\mu = \partial_\mu \left(\sqrt{-g} \,j^\mu\right)$.

Comparing Eqs.~\eqref{eq:Meff} and~\eqref{eq:jeffGW}, we see that the expression in parentheses is a skew 2-index tensor, and can therefore be interpreted as an effective magnetization-polarization tensor.
We can then immediately draw on the results of the previous section, and find
\be
P_i =-h_{i j} E_{j}+\frac{1}{2} h E_{i}+h_{00} E_{i}-\epsilon_{i j k} h_{0 j} B_{k}, \quad
M_i =-h_{i j} B_{j}-\frac{1}{2} h B_{i}+h_{j j} B_{i}+\epsilon_{i j k} h_{0 j} E_{k},
\ee
as stated in Eq.~\eqref{eq:PM}.

As suggested in the main body, an avenue towards the detection of this effective current is repurposing axion haloscopes, where one has a large magnetic field, but $\bE_0 \simeq 0$.
In this case,
\be
\bE_0 \simeq 0:\hspace{0.2cm} P_i =-\epsilon_{i j k} h_{0 j} B_{k}, \quad
M_i =-h_{i j} B_{j}+ \left( h_{j j} -\frac{1}{2} h \right) B_{i}.
\ee
For a general magnetic field configuration, we will have both an effective charge and 3-current, and accordingly corrections to both sourced Maxwell's equations as in Eq.~\eqref{eq:MaxwellPM}.

\subsection{The Biot-Savart Law and Induced Magnetic Fields}

Having discussed general aspects of GW electrodynamics in the previous section, we now turn to the concrete calculations that underpin the detection scheme proposed in the main text.
We focus on an external toroidal magnetic field -- as appropriate for ABRA, SHAFT, and DMRadio-50L -- and compute the induced magnetic field generated by the effective current.

Let us begin by justifying that the appropriate starting point for our calculation is the Biot-Savart law in \eqref{eq:Bz}.
To do so, we show that the effective charge is not required to compute the induced magnetic field to all orders in $\omega$, and that we can neglect the displacement current to ${\cal O}(\omega^4)$.
At ${\cal O}(h)$, we can write Maxwell's equations as
\bea
\nabla \cdot \bE_h &= \rho_\text{eff}, \\
\nabla \cdot \bB_h &= 0, \\
\nabla \times \bE_h &= - \partial_t \bB_h, \\
\nabla \times \bB_h &= \partial_t \bE_h + \jv_\text{eff}.
\label{eq:Maxwellh}
\eea
These equations couple $\bB_h$ and $\bE_h$.
In order to isolate $\bB_h$, we take a curl of the Amp\`ere-Maxwell equation, and then substitute in Faraday's law and the magnetic Gauss' law.
Doing so results in
\be
\Box\, \bB_h = - \nabla \times \jv_\text{eff}.
\label{eq:JefDE}
\ee
From Eq.~\eqref{eq:JefDE}, it is clear that $\bB_h$ can be determined exactly without any reference to $\rho_{\text{eff}}$. (An oscillating charge distribution can of course give rise to a magnetic field, but that can be determined directly from $\jv_\text{eff}$, which is linked to $\rho_\text{eff}$ by the continuity equation.)
A particular solution of Eq.~\eqref{eq:JefDE} assuming there are no boundary surfaces is Jefimenko's equation (see for example Ref.~\cite{jackson_classical_1999})
\be
\bB_h (t, \rv') =  \int \mathrm{d}^3 \rv \,\,  
\left(  \frac{\jv_\text{eff}(t_r,\rv)}{4\pi |\rv'-\rv|^3} 
+\frac{\partial_t \, \jv_\text{eff}  (t_r, \rv)}{4\pi |\rv'-\rv|^2}  
\right)\times (\rv'-\rv),
\hspace{1cm} \text{with}\,\,  t_r=t-|\rv'-\rv|.
\label{eq:Jefimenko}
\ee
The general solution  includes an additional term that depends on the boundary conditions of the instrument. 
We note, however, that determining the effect of the latter  applies for any effective current, both axions and GWs, and this issue has yet to be settled for the conceptually simpler axion problem.
If, at that time, the issue is settled using the language of $\jv_\text{eff}$, then the result can simply be extended to GWs~\footnote{We thank Yoni Kahn for bringing this to our attention.}.

When $1/\omega$ is much larger than the characteristic scale of our instrument -- usually a good approximation when using low-mass axion haloscopes -- we can simplify the result further.
In particular, as the effective current for a plane GW satisfies $\jv_\text{eff}\propto \omega^2 e^{-\i \omega t }$, Eq.~\eqref{eq:Jefimenko} implies that  
\be
\bB_h(t, \rv') =  \int_\text{toroid} \hspace{-0.6cm} \mathrm{d}^3 \rv \,\,  \frac{\jv_\text{eff}(t,\rv) \times (\rv'-\rv)}{4\pi|\rv'-\rv|^3} \left(1+ \frac{1}{2} \omega^2 |\rv'-\rv|^2 +{\cal O}(\omega^3) \right)  
= \int_\text{toroid}\hspace{-0.6cm} \mathrm{d}^3 \rv \,\, \frac{\jv_\text{eff}(t,\rv) \times (\rv'-\rv)}{4\pi|\rv'-\rv|^3} + {\cal O}(\omega^4).
\label{eq:BSlaw}
\ee
Accordingly, if we work only to ${\cal O}(\omega^3)$, it is consistent to neglect displacement currents and start our calculation with the Biot-Savart law.
This is assumed in the main text and will be adopted for remainder of the SM.
That we can neglect the displacement current is equivalent to the magnetoquasistatic approximation conventionally adopted in low-mass axion haloscope calculations, see for example Ref.~\cite{Kahn:2016aff}.

For a pickup loop parallel to the $x-y$ plane, the only relevant component for computing the magnetic flux is 
\begin{eqnarray}
B_z (\rv') &\simeq& \int_\text{toroid}\hspace{-0.6cm}\mathrm{d}^3 \rv \,\,  \frac{  \jv_\text{eff}(t,\rv) \times (\rv'-\rv) \cdot \ez}{4\pi|\rv'-\rv|^3} 
= \int_\text{toroid}\hspace{-0.6cm}\rho\, \mathrm{d}\phi \,\mathrm{d}\rho\,  \mathrm{d}z  \left(\frac{ j_\rho \ephi -j_\phi  \erho }{4\pi|\rv'-\rv|^3}  \right) \cdot (\rv'-\rv), 
\end{eqnarray} 
For a toroidal magnetic field as given in Eq.~\eqref{eq:B0}, the azimuthal and radial components of $\jv_\text{eff}$ can be determined by direct calculation,
\bea
j_{\phi} =& \frac{\omega^2 \bmax R}{\rho} 
\left[ \frac{e^{\kappa}}{\kappa} - \frac{2e^{\kappa}}{\kappa^2} + \frac{2(e^{\kappa}-1)}{\kappa^3} \right]
\left( z\, \hTT_{\rho \phi}\big|_{\rv=0} - \rho\,\hTT_{\phi z}\big|_{\rv=0}  \right)\!, \\
j_\rho =& \frac{\omega^2 \bmax R}{\rho}
\left(
\left[ - \frac{1}{2} - \frac{1}{\kappa} + \frac{2e^{\kappa}}{\kappa^2} + \frac{2(1-e^{\kappa})}{\kappa^3} \right]
\left( \rho\, \hTT_{\rho z}\big|_{\rv=0} + z\,\hTT_{zz}\big|_{\rv=0}  \right) \right.\\
&\hspace{1.8cm}+ \left[ \frac{e^{\kappa}}{\kappa} + \frac{2}{\kappa^2} + \frac{2(1-e^{\kappa})}{\kappa^3} \right]
\left( z\, \hTT_{\rho \rho}\big|_{\rv=0} + z\, \hTT_{zz}\big|_{\rv=0}  \right) \\ 
&\hspace{1.8cm}\left. + \i k_z \left[ \frac{1}{2\kappa} + \frac{1}{2\kappa^2} - \frac{1+2e^{\kappa}}{\kappa^3} + \frac{3(e^{\kappa}-1)}{\kappa^4}
\right] r_i r_j \hTT_{i j}\big|_{\rv=0} \right)\!,
\label{eq:exactj}
\eea
where $\kappa = \i \bk \cdot \rv$.
The components of $\hTT_{ij}$ are determined by, for example, $\hTT_{\rho \rho} = (\erho)_i (\erho)_j \hTT_{ij}$, and take the following explicit forms
\bea
\hTT_{\rho \rho}\big|_{\rv=0} &=
\frac{e^{-\i \omega t}}{\sqrt2} \left(-h^+ (\sin^2(\phi-\phi_h)-\cos^2(\phi-\phi_h)\cos^2 \theta_h ) + 2\,h^\times \cos \theta_h \cos (\phi-\phi_h) \sin (\phi-\phi_h) \right)\!,
\\
\hTT_{\rho \phi}\big|_{\rv=0} &=
\frac{e^{-\i \omega t}}{\sqrt2} \left(-h^+ (1+\cos^2\theta_h) \sin(\phi-\phi_h)\cos(\phi-\phi_h) + h^\times \cos(2(\phi-\phi_h))\cos\theta_h\right)\!,
\\
\hTT_{\rho z}\big|_{\rv=0} &=
-\frac{e^{-\i \omega t}}{\sqrt2} \left(h^+ \cos\theta_h \sin\theta_h \cos(\phi-\phi_h) + h^\times \sin\theta_h \sin(\phi-\phi_h)\right)\!,
\\
\hTT_{\phi z}\big|_{\rv=0} &=
\frac{e^{-\i \omega t}}{\sqrt2} \left(h^+ \cos \theta_h \sin\theta_h \sin(\phi-\phi_h) - h^\times \sin\theta_h \cos(\phi-\phi_h)\right)\!,
\\
\hTT_{zz}\big|_{\rv=0}&=  \frac{e^{-\i \omega t}}{\sqrt2} h^+ \sin^2 \theta_h.
\eea
All expressions within square brackets in Eq.~\eqref{eq:exactj} are ${\cal O}(\omega^0)$ at leading order, and therefore each line, except the last one, contributes at $\omega^2$.
To leading order and taking $\phi_h=0$, these equations reduce to those presented in the main text in Eq.~\eqref{eq:js}.
Similar expressions can also be derived for more complicated forms of the magnetic field, as, for instance, used in the SHAFT experiment.
We will not state the corresponding relations here, although we note that these were accounted for in computing the sensitivity shown in Fig.~\ref{fig:rescale}.

Before moving on to compute the magnetic flux, we note in passing that in our discussion in the main text and in the next subsection, we placed the pickup loop at the vertical center of the apparatus for simplicity.
From the expressions above, we see that this simplifies the discussion as the terms odd in $z$ vanish once we perform the volume integral over the effective current.
However, more general placements of the pickup loop may be of interest, for example to include multiple pickup loops with different geometries, which would allow one to distinguish different incident directions and to differentiate between an axion and a GW signal.
For this purpose, we observe that the terms proportional to $z$ in Eq.~\eqref{eq:js} typically oscillate more strongly with $\phi$ than those proportional to $\rho$, such that the additional signal component which can be recovered away from the vertical center is (mildly) suppressed.
In the case of a single pickup loop, this leads to a mild preference of placing the loop at the vertical center.
We emphasize, however, that the expressions for the effective current and the induced magnetic field given above are fully general and can be used compute the flux through any pickup loop geometry.

\subsection{The Magnetic Flux}

\begin{figure}[!t]
\includegraphics[width=0.4\columnwidth]{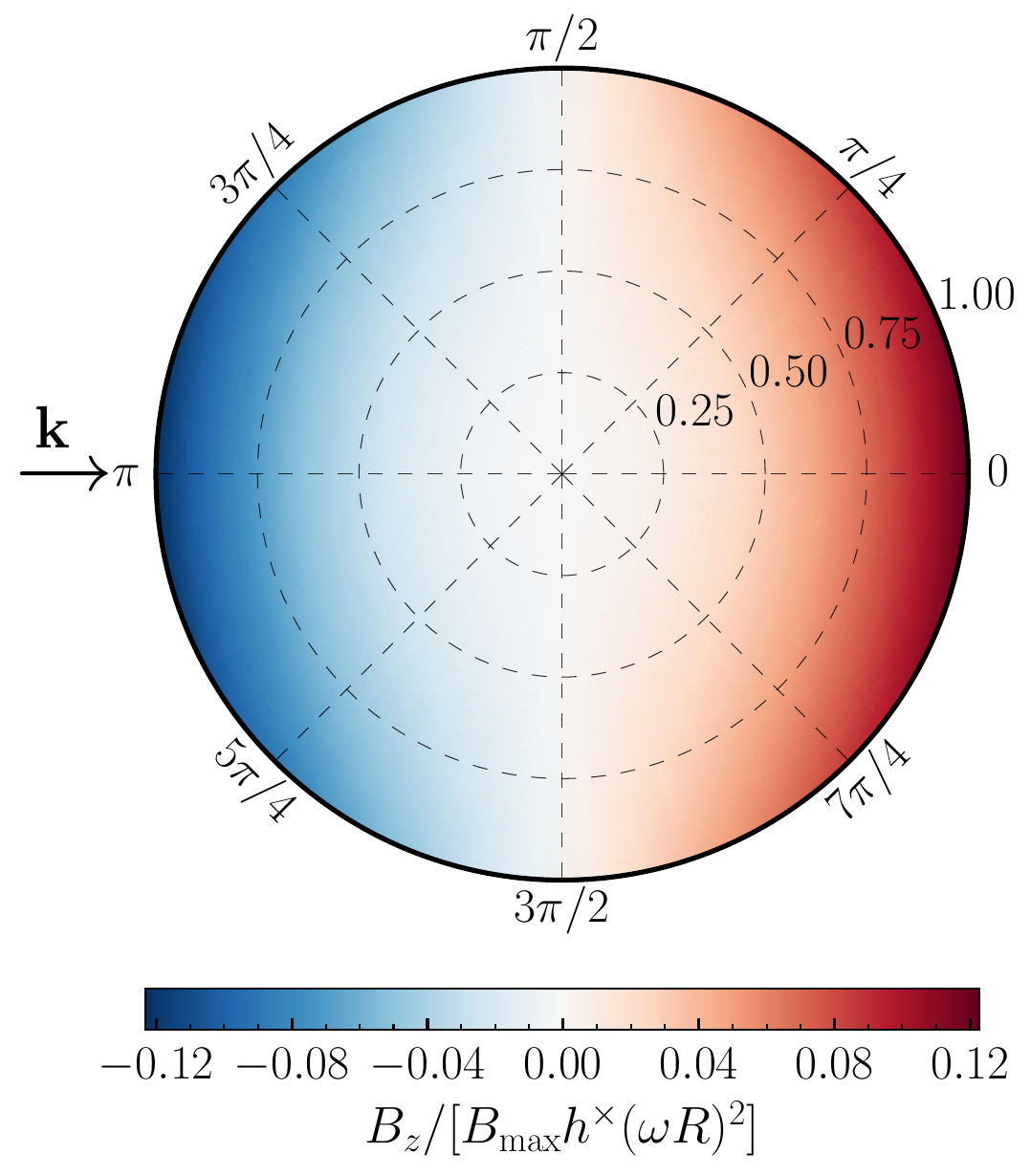}
\caption{The leading $\omega^2$ contribution to the $z$ component of the GW induced magnetic field inside the torus, as a function of the loop coordinates $(\rho',\,\phi')$.
We have explicitly fixed the incident direction of the GW to be $\hat{\bk} = \ex$, as shown.
The sinusoidal variation of the leading magnetic field can clearly be seen, which if integrated over a circular pickup loop will vanish.
Oppositely oriented semicircles -- the figure-8 configuration -- will instead be optimally sensitive to such a field.
For the figure we show only the contribution from the $\times$ polarization, assume an ABRA-10 cm like configuration with $r = R = a = H/4$, show the magnetic field at the vertical center of the torus, and evaluate the $e^{-\i \omega t}$ factor at $t=0$.
In the limit $H \gg R+a$ the quantity we plot can be determined from Eq.~\eqref{eq:leadingBz}.
}
\label{fig:Bz}
\end{figure}

Having calculated the magnetic field, we now turn to the magnetic flux through the pickup loop,
\bea
\Phi = \int_{\text{loop}} \hspace{-0.4cm} \mathrm{d}^2\rv'  B_z (\rv') 
&= \int_{\text{loop}} \hspace{-0.4cm}\rho'\, \mathrm{d}\rho' \mathrm{d}\phi' \int_{\text{toroid}} \hspace{-0.6cm}\rho\, \mathrm{d}\rho\, \mathrm{d}\phi\,  \mathrm{d}z \,\, \frac{ j_\rho \ephi -j_\phi  \erho }{4\pi|\rv'-\rv|^3} \,\cdot (\rv'-\rv).
\label{eq:BSlaw1}
\eea
The integration over the toroid is fixed by the geometry of the magnetic field, and we will restrict our attention to the detector geometry depicted in Fig.~\ref{fig:detector}.
What remains, however, is to fix a configuration for the pickup loop.
As in the main text, we will consider two cases.
In order to highlight the distinction, in Fig.~\ref{fig:Bz} we plot the leading contribution to $B_z (\rv')$ for a GW incident along the $x$-direction, taking $r=R=a=H/4$.
The sinusoidal variation emphasized from the simplified result in Eq.~\eqref{eq:leadingBz} is clearly visible.
From this, it follows that a circular pickup loop will be insensitive to the leading flux---the contribution from opposite sides of the pickup loop are cancelled.
Nevertheless, as this is the loop geometry used by axion experiments we will consider it below.
We will then consider the case of a partial loop, from which the optimal readout of a figure-8 geometry, where the contributions from each side can be added coherently, can be inferred.

\emph{Circular loop:}
This is a highly symmetric configuration, allowing significant simplification.
Let us first notice that $\phi'$ in Eq.~\eqref{eq:BSlaw1} only enters in the combination $\rv'-\rv$: it does not appear in the effective current.
Explicitly, we have
\be
\rv'-\rv = -\left(\rho-\rho' \cos(\phi'-\phi)\right) \,\erho +\rho' \sin(\phi'-\phi)\,\ephi-z\,\ez,
\quad
|\rv'-\rv|^2 = z^2+\rho^2+\rho'^2-2\rho\rho' \cos(\phi'-\phi).
\ee
In particular, we observe that $\phi'$ only appears in the combination $\phi'-\phi$.
We can thus largely remove $\phi$ by performing a shift $\phi'\to \phi'+\phi$, leaving
\be
\Phi = 
\frac{1}{4\pi} \int_{\text{toroid}} \hspace{-0.6cm}\rho\,\mathrm{d}\rho\,\mathrm{d}\phi\, \mathrm{d}z 
\int_{\text{loop}}  \hspace{-0.4cm}\rho' \mathrm{d}\rho'\mathrm{d}\phi'\,
\frac{j_\rho \rho' \sin\phi' +j_\phi (\rho-\rho' \cos\phi')}{(z^2+\rho^2+\rho'^2-2\rho\rho' \cos \phi')^{3/2}}.
\ee

The shift will also impact the range over which $\phi'$ is integrated.
If, however, we integrate around a complete circle (as required for a circular pickup loop), then this shift has no impact: the range remains $\phi' \in [0,2\pi)$.
There are then two simplifications that occur.
Firstly, the term proportional to $j_{\rho}$ vanishes as can be confirmed by performing the $\phi'$ integral (note, for example, that the integrand is odd in $\phi'$).
This is true independent of the form of $j_{\rho}$, as mentioned in the main text.
Secondly, the only $\phi$ dependence left in the integrand sits in $j_{\phi}$, so that we can rewrite the flux as
\be
\Phi = 
\frac{1}{4\pi} \int_{\text{toroid}} \hspace{-0.6cm}\rho\,\mathrm{d}\rho\,\mathrm{d}z 
\int_{\text{loop}}  \hspace{-0.4cm}\rho' \mathrm{d}\rho'\mathrm{d}\phi'\,
\left[ \frac{\rho-\rho' \cos\phi'}{(z^2+\rho^2+\rho'^2-2\rho\rho' \cos \phi')^{3/2}} \right] \left[ \int^{2\pi}_0 \mathrm{d}\phi\, j_\phi \right]\!.
\label{eq:Flux1}
\ee
Using the form of $j_{\phi}$ in Eq.~\eqref{eq:exactj} we can perform the integral over $\phi$ to obtain
\be
\int^{2\pi}_0 \mathrm{d}\phi\, j_\phi  
= \i\sqrt{2}  \pi\omega h^\times \bmax\frac{R}{\rho} \,J_2 (\omega \rho \sin \theta_h)  e^{\i \omega (z \cos \theta_h-t)}
=  \left(\frac{\pi\omega^3 \rho \sin^2 \theta_h\bmax\, R}{4\sqrt2} \right)\, \i h^\times e^{-\i \omega t}   +{\cal O}(\omega^4).
\label{eq:w3PhiAp}
\ee
As in Eq.~\eqref{eq:jphiint} in the main text, this demonstrates that the $+$ polarization does not generate any signal in a circular pickup loop, and that the leading contribution is ${\cal O}(\omega^3)$.

In the main text we presented results derived assuming $R,a\ll H\ll 1/\omega$.
In this limit,  $e^{\i \omega z \cos \theta_h}\simeq 1$, the terms in Eq.~\eqref{eq:Flux1} odd in $z$ vanish for a pickup loop placed at the vertical center of the apparatus, and the only remaining integral over $z$ to perform is
\be
\int^{H/2}_{-H/2} \frac{\mathrm{d}z}{(z^2+|\rv'-\rv|_{z=0}^2)^{3/2}} = \frac{2}{|\rv'-\rv|_{z=0}^2} \left[1+ {\cal O}{\left(|\rv'-\rv|_{z=0}^2/H^2\right)}\right]\!.
\ee
Together with Eq.~\eqref{eq:w3PhiAp}, this leads to Eq.~\eqref{eq:w3Phi} in the main text.

\emph{Partial loop:}
As shown by the calculation leading to \eqref{eq:w3PhiAp}, the decoupling of $h^+$ together with the absence of an ${\cal O}(\omega^2)$ contribution is a consequence of the reflection symmetry of the circular pickup loop.
Motivated by this, we consider a pickup loop forming a circular arc subtending the angles $0<\phi'<\phi'_\text{max}$.
As explained above, the integration along the $z$-axis may be formally performed by replacing $ dz/|\rv'-\rv|^3\to 2/|\rv'-\rv|_{z=0}^2\,$ and setting $z=0$ everywhere else.
Under the this approximation, the flux in Eq.~\eqref{eq:BSlaw1} takes the form
\bea
\Phi &= \int^{R+a}_{R} \,\mathrm{d}\rho\,  \rho  \int^{2\pi}_{0}  \mathrm{d}\phi\int^r_0 \mathrm{d}\rho' \rho' \int^{\phi'_\text{max}}_{0} \mathrm{d}\phi' \,\, \left[ \frac{ j_\rho \ephi -j_\phi  \erho }{2\pi|\rv'-\rv|^2} \,\cdot (\rv'-\rv) \right]_{z=0}\\
&=
\frac{e^{-\i \omega t}}{12 \sqrt{2}} \omega^2 \bmax  r^3 R \ln (1+a/R) \sin \theta_h \\
&\times \left( -h^+ \cos \theta_h [\cos \phi_h-\cos (\phi'_\text{max}-\phi_h) ]
+ h^\times [\sin\phi_h+\sin (\phi'_\text{max}-\phi_h)] \right)\!.
\eea
The flux for specific geometries can be immediately obtained from this result.
This includes the figure-8 configuration discussed in the main text; in particular, Eq.~\eqref{eq:w2Phi} follows from this result.

\section{Gravitational Wave Sources in the Ultra-High Frequency Band}
\label{sec:sources}

Here we summarize the possible sources that the search proposed in the main text may detect, with a particular focus on the case of PBH binaries.
The discussion is not intended to be exhaustive, and we refer to Ref.~\cite{Aggarwal:2020olq}.
There are no known sizable astrophysical sources in this frequency range, implying that any GW detection at UHF is a smoking gun signal of new physics.
Many models that posit a completion of the Standard Model of particle physics at high energies also predict additional dynamics in the early Universe (such as phase transitions, formation of topological defects, or non-perturbative (p)reheating dynamics after inflation) which source gravitational radiation that can, depending on model parameters, constitute a sizable fraction of the total radiation energy present in the early Universe.
Taking into account the cosmological red-shift, these stochastic GWs are observable at frequencies $f \sim 100~\text{MHz}/\epsilon_* \, (T_*/10^{15}~\text{GeV})$ where $T_*$ is the temperature of the Universe when the GWs are sourced and $\epsilon_* < 1$ indicates the GW wavelength in units of the Hubble horizon at the time of production.
This result suggests that the UHF band could be ideal for searching for new dynamics present at energies well beyond what we can hope to probe directly.
However, such cosmological GWs contribute to the radiation energy budget of the Universe and thus impact big bang nucleosynthesis and the decoupling of the cosmic microwave background.
They are therefore constrained by bounds on the effective number of additional neutrinos, $\Delta N_\text{eff}$~\cite{Pisanti:2020efz,Yeh:2020mgl}, in particular $\rho_\text{GW}/\rho_c \lesssim 10^{-5} \Delta N_\text{eff}$, where $\rho_c$ is it critical density.
For a broadband spectrum, this constrains the characteristic strain to be at most $h_{c,\text{sto}} \lesssim 10^{-29} \, (100~\text{MHz}/f) \, \Delta N_\text{eff}^{1/2}$.
This is several orders of magnitude below the most optimistic sensitivities we presented in Fig.~\ref{fig:projectedlimits}, suggesting that even with the advantageous volume scaling of the detection strategy suggested in this work, such a signal is out of reach in the foreseeable future.

Such strong constraints do not apply to isolated GW sources in the late Universe.
To highlight this, we discuss in more detail the possible signal from PBH binaries, before briefly commenting on other possible sources in the late Universe at the end of this appendix.
For simplicity, let us consider a circular binary of two PBHs which have equal mass $m_\PBH$. 
Then, the amplitude of the GW signal emitted from a binary at a distance $D$ along the symmetry axis of the circular orbit is~\cite{Maggiore:2007ulw}
\be
h_{+,\times}^\PBH(f, m_\PBH, D) \simeq  1.3 \times 10^{-23} \left( \frac{10~\text{kpc}}{D}\right)  \left(\frac{m_\PBH}{10^{-5} M_\odot}\right)^{5/3} \left( \frac{ f}{100~\text{MHz}} \right)^{2/3}\!.
\label{eq:hBinary}
\ee
As time evolves, GW emission leads the binary to lose energy, and consequently the GW amplitude and frequency increase until the merger, at which time
\be
f \simeq 220~\text{MHz} \, \left(\frac{10^{-5} M_\odot}{m_\PBH} \right) \!.
\label{eq:fISCO}
\ee
We emphasize that from the above it can be seen that frequencies in the MHz band and beyond correspond to light black holes, with $m_\PBH \ll M_\odot$, which excludes a stellar-origin for the black hole sources in this frequency band.

The probability of observing such a PBH binary is determined by the formation rate of the binaries and their local density. 
The light PBHs we are interested in are dominantly formed in the early Universe.
Through the distribution of all PBHs, a subset will be close enough that their gravitational attraction leads the pair to decouple from the Hubble flow and form a gravitational bound object.
(In this scenario, a third PBH is in fact required in the process to ensure the conservation of angular momentum.)
Assuming a Gaussian distribution for the primordial density perturbations, normalized to give a PBH abundance which constitutes a fraction $f_\PBH$ of the total dark-matter density, PBHs are formed in rare events following Poisson statistics and the binary formation rate is obtained as~\cite{Raidal:2018bbj,Hutsi:2020sol}
\bea
R_0(m_\PBH,f_\PBH) &\simeq 6.6 \times 10^{-8}\,\text{kpc}^{-3}\,\text{yr}^{-1} \, f_\PBH^{53/37} \left( \frac{m_\PBH}{10^{-5} M_\odot} \right)^{-32/37} S_\text{early}(f_\PBH)\, S_\text{late}(f_\PBH), \\
S_\text{early}(f_\PBH) &=\text{min}\left\{1,\left(\frac{f_\PBH}{0.01}\right)^{1/2}\right\}\!, \\
S_\text{late}(f_\PBH) &= \text{min}\left\{1,9.6 \times 10^{-3} f_\PBH^{-0.65} e^{0.03\ln^2 f_\PBH}\right\}\!,
\label{eq:R0}
\eea
where for simplicity we have taken the PBH mass distribution to be concentrated at $m_\PBH$.
This formation rate receives corrections from different effects, see Sec. 4.6 of Ref.~\cite{Franciolini:2021nvv} for a detailed summary and relevant references. 
We include two effects which suppress the merger rate, denoted by $S_\text{early}$ and $S_\text{late}$ in Eq.~\eqref{eq:R0}.
The first of these is a backreaction from the surrounding matter in the early Universe, which suppresses the merger rate for $f_\PBH < 0.01$~\cite{Hutsi:2020sol}.
Since microlensing bounds largely constrain $f_\PBH$ to lie below this value~\cite{Carr:2021bzv}, this effectively changes the exponent of $f_\PBH$ in Eq.~\eqref{eq:R0} from 53/37 to $\sim$2.
(For an alternative treatment of this effect, see Refs.~\cite{Hutsi:2020sol,Franciolini:2022xyz}, although we confirmed adopting this alternative form for $S_\text{early}$ does not significantly alter our results.)
Interactions of the binary with matter in the late Universe can also disrupt the system, again suppressing the merger rate, as encoded in $S_\text{late}$~\cite{Hutsi:2020sol}.
It is also expected that accretion will impact the merger rate, however the degree to which it does so is less certain, and so for the simple estimate we provide here, we neglect the effect of accretion.
We note, however, that a recent study in Ref.~\cite{DeLuca:2020bjf} suggested that for the PBH masses we consider, significant accretion is unlikely, so that neglecting its effect may well be a reasonable assumption.

\begin{figure}
\centering
\includegraphics[width = 0.5 \textwidth]{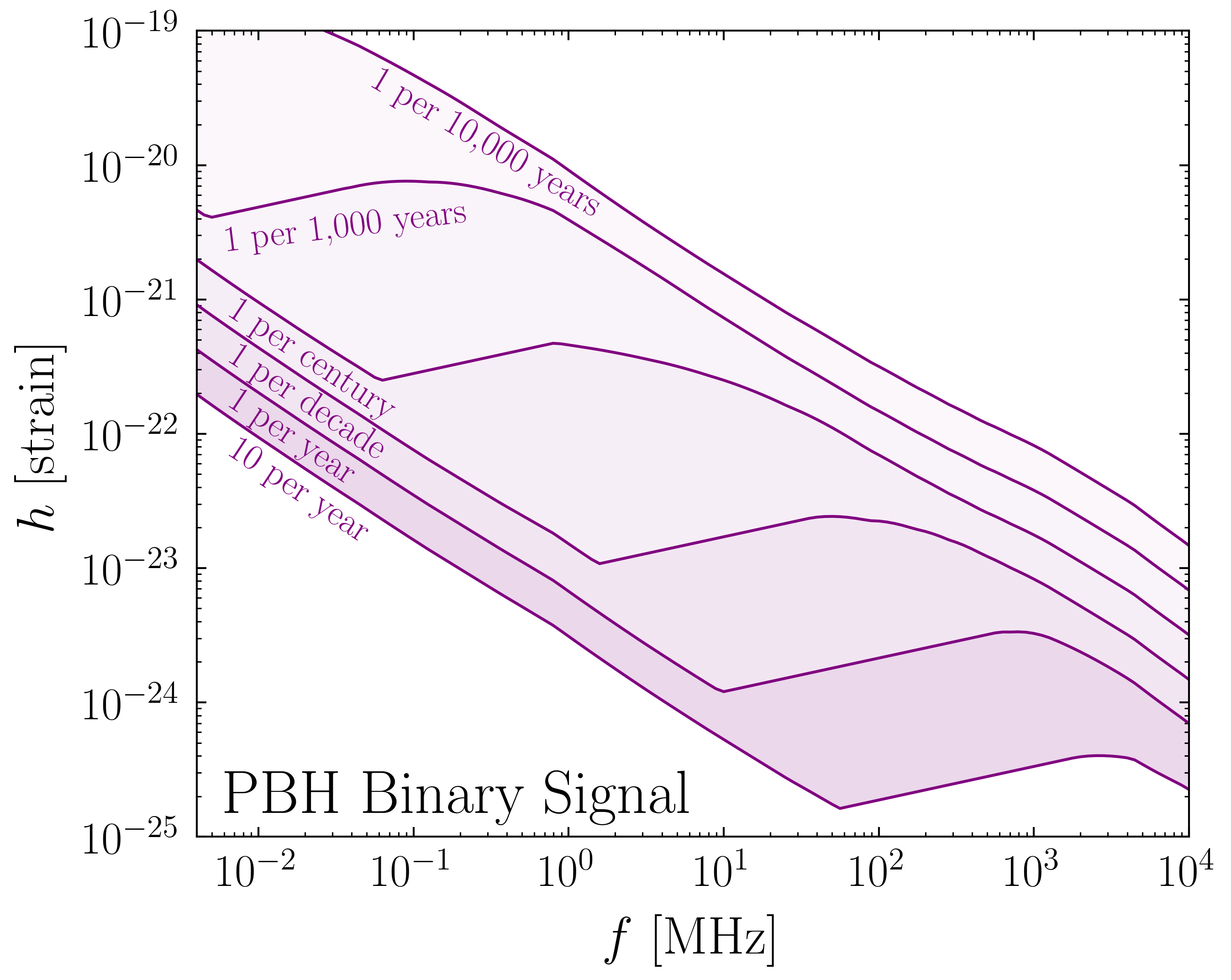}
\caption{Strain sensitivity $h_\text{th}$ required to observe a signal from PBH binary systems (either from the inspiral or merger phase) at rates between ten events per year and one per ten-thousand years.
The case for one per year was reproduced in Fig.~\ref{fig:projectedlimits}.
Details of how these sensitivities were computed are provided in the text, however, we caution that there remain uncertainties in this computation (for instance from the impact of accretion on the merger rate) that we have neglected.
}
\label{fig:PBH}
\end{figure}

From Eq.~\eqref{eq:R0}, we obtain the number of expected merger events per year in a volume of $1~\text{kpc}^3$, for a given $m_\PBH$ and $f_\PBH$.
If we have a GW telescope with a strain sensitivity $h_\text{th}$ at a frequency $f$, then using Eq.~\eqref{eq:hBinary}, we can determine what distance to which we can detect such a binary signal, for a given $m_\PBH$.
Combining the two results, we can determine the expected rate at which the telescope will observe a signal from PBH binary events,
\be
\langle \Gamma \rangle = \int_0^\infty dr\, 4 \pi r^2 \delta(r) \,  R_0(m_\PBH, f_\PBH) \,  \Theta \left[ Q^{1/4} \,  h_{+,\times}^\PBH(f, m_\PBH, r) - h_\text{th} \right]\!.
\label{eq:Gamma}
\ee
There are two additional features we have added to this expression beyond the above discussion.
Firstly, the quality factor or coherence of the signal, $Q$, is proportional to the time remaining until the merger given an emitted GW frequency $f$ and PBH mass $m_\PBH$, yielding $Q \propto f^2/\dot f \propto m_\PBH^{-5/3}$~\cite{Maggiore:2007ulw}.
For a binary system very close to merger, $\dot f/f^2 \sim 1$ and hence $Q \sim 1$. 
Consequently, we can normalize $Q$ to the mass which would be at the point of merger for the frequency considered, using Eq.~\eqref{eq:fISCO}.
As discussed in the main text, coherent signals are more readily detected, and so we have included this factor as an enhancement of $Q^{1/4}$.
Note that the scaling $Q^{1/4}$ is specific to the detector concept proposed here. When comparing to generic GW detectors, one often instead uses the characteristic strain, $h_c = f \, \tilde h(f)$ with $\tilde h(f)$ denoting the Fourier transform of the GW signal. For a PBH binary sufficiently far from merger,  $h_{c} \simeq Q^{1/2} \, h_{+, \times}^\text{PBH}$.
We caution that this caveat also applies when comparing to some of the detector sensitivities shown in grey in Fig.~\ref{fig:projectedlimits}.
(More generally, the distinction between $Q^{1/2}$ and $Q^{1/4}$ scaling for axion haloscopes depends on whether the observation time is less than or greater than the signal coherence time, and we have assumed the latter throughout.)

Secondly, $R_0$ is the rate averaged over the cosmological volume of the binaries, whereas locally we live in a significant matter overdensity, the Milky Way.
This reality is included in the factor of $\delta(r)$.
We assume that the PBH binaries simply track the overall DM abundance, resulting in a simple linear dependence on $\delta(r)$ in the integrand. It enters in particular with a different power than a primordial overdensity or $f_\PBH$ enter, since only the latter impact the binary formation processes in the early Universe.
As pointed out in Ref.~\cite{Pujolas:2021yaw}, for light PBHs the local DM overdensity in the Milky Way halo can boost the observed merger rate significantly.
We parametrize the Milky Way halo by a NFW mass profile~\cite{Navarro:1995iw,Navarro:1996gj} up to the virial radius $r_{200} \simeq 207.19$~kpc, with a local density $\rho_\odot \simeq 0.31 \text{GeV/cm}^3$ at the location of the solar system, $r_\odot \simeq 8.13$~kpc~\cite{GRAVITY:2018ofz}, where the halo parameters are determined from Ref.~\cite{Cautun:2019eaf}.

Fig.~\ref{fig:PBH} shows the detector sensitivity required to detect a fixed expected number of merger event per year, one curve of which was included in Fig.~\ref{fig:projectedlimits}.
A relatively large event rate, $\langle \Gamma \rangle = 10$/year requires a very good detector sensitivity, whereas rare event may be detected with a more moderate detector sensitivity.
Here we have marginalized over the PBH mass in Eq.~\eqref{eq:Gamma}, i.e.\ for any given frequency we chosen $m_\PBH$ such as to maximize the rate $\Gamma$. 
The fraction of PBH dark matter $f_\PBH$ is taken to saturate the microlensing bound at that PBH mass~\cite{Carr:2021bzv}.
Due to the mild scaling with $Q$,   this rate is dominated by merger events rather than the early inspiral phase in most of the parameter space.
For fixed values of $m_\text{PBH}$ and $f_\text{PBH}$ these results are consistent with Ref.~\cite{Franciolini:2022xyz} after recasting the dimensionless GW amplitude $h$ in terms of the characteristic strain $h_c$.
(Note that such high-frequency GW bursts can also be constrained by the non-observation of corresponding GW memory signals in ground-based interferometers~\cite{McNeill:2017uvq}; see also Ref.~\cite{Lasky:2021naa}.)
In regions where the density profile of the Milky Way halo becomes irrelevant because the effective detector volume set by the Heaviside function in Eq.~\eqref{eq:Gamma} is much larger (at small frequencies) than the Milky Way halo, the strain sensitivity scales as $h \propto m_\PBH^{79/111} \propto f^{-79/111}$, where the last relation only holds for signals close to merger, see Eq.~\eqref{eq:fISCO}, and we have neglected $Q$.
The difference in overall normalization between these two regions is given by the local DM overdensity at the position of the solar system. When the effective detector volume becomes comparable to the size of the halo, the DM density profile leads to the feature visible at intermediate frequencies in Fig.~\ref{fig:PBH}.

We end with several brief comments on additional late time sources.
PBH can furthermore source GWs when they scatter off each other without merging.
The resulting spectrum is not monochromatic and peaks at a frequency determined by the specific hyperbolic orbit that the PBHs follow.
Observable signals require significant clustering or very rare events such as close to parabolic encounters~\cite{Garcia-Bellido:2017knh,Garcia-Bellido:2021jlq,Morras:2021atg}.
Another possible late time source is axion superradiance~\cite{Brito:2015oca}. 
Clouds of axion-like particles could form around black holes when the axion Compton wavelength matches the Schwarzschild radius.
This would lead to an emission of GWs with a wavelength set by this same scale,
\be
\frac{f}{100~\text{MHz}} \sim  \frac{m_a}{10^{-7}~\text{eV}} \sim  \frac{10^{-3} M_\odot}{m_\text{BH}}.
\ee
Sourcing GWs with a frequency above 0.1~MHz via this mechanism thus requires primordial black holes with masses below the Chandrasekhar limit, $M_\text{BH} < 1.4  M_\odot$. The amplitude of GWs originating from axion decay can then be estimated as
\be
h \sim 10^{-26}\, \frac{100~\text{MHz}}{f} \frac{10~\text{kpc}}{D},
\ee
assuming that the axion cloud constitutes $0.1 \%$ of the black hole mass.
The signal is expected to be highly monochromatic.

These examples illustrate that different search strategies will need to be implemented to optimally search for different possible GW sources.
Generally, however, the GW signal is expected to be less coherent than axion dark-matter, and so in general strategies searching for broader signals in the frequency domain will need to be devised.
(In this sense there are strong similarities with the search for relativistic axions, for further discussion see Ref.~\cite{Dror:2021nyr}.)

\end{document}